\documentclass[nonacm, dvipsnames, sigconf]{acmart}

\usepackage{graphicx}
\usepackage{listings}
\usepackage{hyperref}
\usepackage{dblfloatfix}
\usepackage{tikz-cd}
\usepackage{bm}
\usepackage{listings}
\usepackage{mathtools}
\usepackage{mathpartir}
\usepackage{float}
\usepackage{subcaption}
\usepackage{xcolor}
\usepackage[safe]{tipa}
\usepackage[inline]{enumitem}

\usepackage{xcolor}
\usepackage{forloop}
\usepackage{ifthen}
\usepackage{xifthen}
\usepackage{calc}
\usepackage{xspace}
\usepackage{mathrsfs}

\newcounter{i}
\newcounter{j}
\newcounter{n1}

\newcommand{\bnfdef}{\ensuremath{\Coloneqq}}
\newcommand{\bnfalt}{\ensuremath{\mid}\xspace}

\renewcommand{\epsilon}{\varepsilon}

\newcommand{\partialfunc}{\rightharpoonup}

\newcommand{\derive}[3][]{%
    \ifthenelse{\isempty{#1}}{%
        \frac{\text{d} #2}{\text{d} #3}%
    }{%
        \frac{\text{d}^{#1} #2}{\text{d} #3^{#1}}%
    }%
}

\newcommand{\true}{\textbf{\textsf{T}}}
\newcommand{\false}{\textbf{\textsf{F}}}

\newcommand{\Comment}[1]{}
\newcommand{\Space}[1]{}

\ifdefined\boundary\else\newcommand{\boundary}{\ensuremath{\partial}}\fi

\newcommand{\proves}{\vdash}

\ifdefined\theorem\else\newtheorem{theorem}{Theorem}\fi
\ifdefined\lemma\else\newtheorem{lemma}{Lemma}\fi
\ifdefined\proposition\else\newtheorem{proposition}{Proposition}\fi
\ifdefined\claim\else\newtheorem{claim}{Claim}\fi
\ifdefined\corollary\else\newtheorem{corollary}{Corollary}\fi
\ifdefined\definition\else\newtheorem{definition}{Definition}\fi
\ifdefined\example\else\newtheorem{example}{Example}\fi
\ifdefined\remark\else\newtheorem{remark}{Remark}\fi
\ifdefined\notation\else\newtheorem{notation}{Notation}\fi

\newcommand{\N}{\ensuremath{\mathbb{N}}}

\newcommand{\id}[1]{\ensuremath{\bm{1}_{#1}}}

\newcommand{\Sym}[1]{\ensuremath{\text{Sym}\parens{#1}}}

\newcommand{\notationSym}[1][X]{Let $\Sym{X}$ be the group of bijections of $X$.\xspace}

\newcommand{\brackets}[3]{\ensuremath{{\left#1 {#3} \right#2}}}
\newcommand{\parens}[1]{\brackets{(}{)}{#1}}
\newcommand{\angles}[1]{\brackets{<}{>}{#1}}
\newcommand{\curlys}[1]{\brackets{\{}{\}}{#1}}

\newcommand{\tif}{\ensuremath{~\text{if}~}}
\newcommand{\tsuchthat}{\ensuremath{~\text{s.t.}~}}

\newcommand{\diagmatrix}[2]{%
    \setcounter{n1}{#1 - 1}
    \begin{pmatrix}
        \forloop{i}{0}{\value{i} < #1}{%
            \forloop{j}{0}{\value{j} < #1}{%
                \ifthenelse{\equal{\value{i}}{\value{j}}}{#2}{0}
                \ifthenelse{\value{j} < \value{n1}}{&}{}
            }
            \ifthenelse{\value{i} < \value{n1}}{\\}{}
        }
    \end{pmatrix}
}

\newcommand{\finitecommabody}[1]{}
\newcommand{\countablecommabody}[1]{}

\newcommand{\deffuncparam}[6]{%
    \expandafter\newcommand\csname domain#1\endcsname{\ensuremath{#3}}%
    \expandafter\newcommand\csname codomain#1\endcsname{\ensuremath{#5}}%
    \expandafter\newcommand\csname decl#1\endcsname[1]{\ensuremath{#2_{##1} : #3 #4 #5}}%
    \expandafter\newcommand\csname funcname#1\endcsname[1]{\ensuremath{#2_{##1}}}%
    \expandafter\newcommand\csname body#1\endcsname[2]{\ensuremath{#6}}%
    \expandafter\newcommand\csname call#1\endcsname[2]{\ensuremath{#2_{##1} \parens{##2}}}%
    \expandafter\newcommand\csname callbody#1\endcsname[2]{\ensuremath{\expandafter\csname call#1\endcsname{##1}{##2} = \expandafter\csname body#1\endcsname{##1}{##2}}}%
    \expandafter\newcommand\csname map#1\endcsname[2]{\ensuremath{##2 \mapsto \expandafter\csname body#1\endcsname{##1}{##2} }}%
    \expandafter\newcommand\csname inversecall#1\endcsname[2]{\ensuremath{#2_{##1} \parens{##2}^{-1}}}%
    \expandafter\newcommand\csname image#1\endcsname[1]{\ensuremath{\text{im}\parens{#2_{##1}}}}%
}

\newcommand{\defbinfunc}[6]{%
    \expandafter\newcommand\csname domain#1\endcsname{\ensuremath{#3}}%
    \expandafter\newcommand\csname codomain#1\endcsname{\ensuremath{#5}}%
    \expandafter\newcommand\csname decl#1\endcsname{\ensuremath{#2 : #3 #4 #5}}%
    \expandafter\newcommand\csname funcname#1\endcsname{\ensuremath{#2}}%
    \expandafter\newcommand\csname body#1\endcsname[2]{#6}%
    \expandafter\newcommand\csname call#1\endcsname[2]{\ensuremath{#2 \parens{##1, ##2}}}%
    \expandafter\newcommand\csname callpair#1\endcsname[1]{\ensuremath{#2 \parens{##1}}}%
    \expandafter\newcommand\csname callbody#1\endcsname[2]{\ensuremath{\expandafter\csname call#1\endcsname{##1}{##2} = \expandafter\csname body#1\endcsname{##1}{##2}}}%
    \expandafter\newcommand\csname map#1\endcsname[2]{\ensuremath{\parens{##1, ##2} \mapsto \expandafter\csname body#1\endcsname{##1}{##2} }}%
}

\newcommand{\defbinop}[6]{%
    \expandafter\newcommand\csname domain#1\endcsname{\ensuremath{#3}}%
    \expandafter\newcommand\csname codomain#1\endcsname{\ensuremath{#5}}%
    \expandafter\newcommand\csname decl#1\endcsname{\ensuremath{#2 : #3 #4 #5}}%
    \expandafter\newcommand\csname funcname#1\endcsname{\ensuremath{#2}}%
    \expandafter\newcommand\csname body#1\endcsname[2]{#6}%
    \expandafter\newcommand\csname call#1\endcsname[2]{\ensuremath{##1 #2 ##2}}%
    \expandafter\newcommand\csname callbody#1\endcsname[2]{\ensuremath{\expandafter\csname call#1\endcsname{##1}{##2} = \expandafter\csname body#1\endcsname{##1}{##2}}}%
    \expandafter\newcommand\csname map#1\endcsname[2]{\ensuremath{\parens{##1, ##2} \mapsto \expandafter\csname body#1\endcsname{##1}{##2} }}%
}

\newcommand{\deffunc}[6]{%
    \expandafter\newcommand\csname domain#1\endcsname{\ensuremath{#3}}%
    \expandafter\newcommand\csname codomain#1\endcsname{\ensuremath{#5}}%
    \expandafter\newcommand\csname decl#1\endcsname{\ensuremath{#2 : #3 #4 #5}}%
    \expandafter\newcommand\csname funcname#1\endcsname{\ensuremath{#2}}%
    \expandafter\newcommand\csname body#1\endcsname[1]{#6}%
    \expandafter\newcommand\csname call#1\endcsname[1]{\ensuremath{#2 \parens{##1}}}%
    \expandafter\newcommand\csname callbody#1\endcsname[1]{\ensuremath{\expandafter\csname call#1\endcsname{##1} = \expandafter\csname body#1\endcsname{##1}}}%
    \expandafter\newcommand\csname map#1\endcsname[1]{\ensuremath{##1 \mapsto \expandafter\csname body#1\endcsname{##1} }}%
    \expandafter\newcommand\csname invcall#1\endcsname[1]{\ensuremath{#2^{-1}\parens{##1}}}%
    \expandafter\newcommand\csname image#1\endcsname{\ensuremath{\text{im}\parens{#2}}}%
}

\usepackage{listings, xcolor}

\definecolor{verylightgray}{rgb}{.97,.97,.97}

\lstdefinelanguage{Solidity}{
	keywords=[1]{anonymous, assembly, balance, call, callcode, class, constant, constructor, contract, debugger, delegatecall, delete, emit, event, experimental, export, external, function, gas, implements, import, in, indexed, instanceof, interface, internal, is, length, library, log0, log1, log2, log3, log4, memory, modifier, new, payable, pragma, private, protected, public, pure, push, return, returns, selfdestruct, send, solidity, storage, struct, suicide, super, then, throw, typeof, using, view, with, addmod, ecrecover, keccak256, mulmod, ripemd160, sha256, sha3}, 
	keywordstyle=[1]\color{Rhodamine}\bfseries,
	keywords=[2]{address, bool, byte, bytes, bytes1, bytes2, bytes3, bytes4, bytes5, bytes6, bytes7, bytes8, bytes9, bytes10, bytes11, bytes12, bytes13, bytes14, bytes15, bytes16, bytes17, bytes18, bytes19, bytes20, bytes21, bytes22, bytes23, bytes24, bytes25, bytes26, bytes27, bytes28, bytes29, bytes30, bytes31, bytes32, enum, int, int8, int16, int24, int32, int40, int48, int56, int64, int72, int80, int88, int96, int104, int112, int120, int128, int136, int144, int152, int160, int168, int176, int184, int192, int200, int208, int216, int224, int232, int240, int248, int256, mapping, string, uint, uint8, uint16, uint24, uint32, uint40, uint48, uint56, uint64, uint72, uint80, uint88, uint96, uint104, uint112, uint120, uint128, uint136, uint144, uint152, uint160, uint168, uint176, uint184, uint192, uint200, uint208, uint216, uint224, uint232, uint240, uint248, uint256, var, void, ether, finney, szabo, wei, days, hours, minutes, seconds, weeks, years},	
	keywordstyle=[2]\color{Emerald}\itshape,
	keywords=[3]{block, balance, blockhash, coinbase, difficulty, gaslimit, number, timestamp, msg, data, gas, sender, sig, now, tx, gasprice, origin, this},	
	keywordstyle=[3]\color{Lavender}\itshape,
	keywords=[4]{assert, break, case, catch, continue, default, else, false, finally, for, if, require, revert, switch, throw, true, try, while, do}, 
	keywordstyle=[4]\color{Cerulean}\bfseries,
	identifierstyle=\color{black},
	sensitive=false,
    keepspaces=true,
    columns=fullflexible,
	comment=[l]{//},
	morecomment=[s]{/*}{*/},
    commentstyle=\color{CadetBlue}\textit,
	stringstyle=\color{ForestGreen},
	morestring=[b]',
	morestring=[b]",
    xleftmargin=5.0ex,
    extendedchars=true,
	basicstyle=\small\ttfamily,
	showstringspaces=false,
	showspaces=false,
	tabsize=2,
	breaklines=true,
	showtabs=false,
	captionpos=b
}

\definecolor{verylightgray}{rgb}{.97,.97,.97}

\lstdefinelanguage{flow}{
    keywords=[1]{var,interface,where,contract,transaction,view,transformer,type,is,returns},
    keywordstyle=[1]\color{Rhodamine}\bfseries,
    keywords=[2]{false,true,sometimes,fungible,asset,unique,immutable,consumable},
    keywordstyle=[2]\color{DarkOrchid}\bfseries,
    keywords=[3]{map,linking,set,list,address,uint256,record,table,string,bytes,nat,bool},
    keywordstyle=[3]\color{Emerald}\itshape,
    keywords=[4]{nonempty,any,every,empty,exactly,one,msg,this,in,new,or,and,not,total,consume},
    keywordstyle=[4]\color{Lavender}\itshape,
    keywords=[5]{if,else,only,when,try,catch,such,st,that},
    keywordstyle=[5]\color{Cerulean}\bfseries,
    keepspaces=true,
    sensitive=false, 
    morecomment=[l]{//}, 
    morecomment=[s]{/*}{*/}, 
    commentstyle=\color{CadetBlue}\textit,
    morestring=[b]", 
    stringstyle=\color{ForestGreen}, 
    showstringspaces=false,
    xleftmargin=5.0ex,
    mathescape=true,
    escapechar=|,
    basicstyle=\small\ttfamily
}

\lstnewenvironment{flow}
{%
    \lstset{language=flow, basicstyle=\small\ttfamily, mathescape=true}%
}
{
}

\newcommand{\flowinline}[1]{\lstinline[language=flow,basicstyle=\small\ttfamily,mathescape]{#1}}

\renewcommand{\true}{\textbf{\texttt{true}}\xspace}
\renewcommand{\false}{\textbf{\texttt{false}}\xspace}

\renewcommand{\implies}{\Rightarrow}

\newcommand{\with}{\text{with}\xspace}

\newcommand{\Loc}{\mathcal{L}}
\newcommand{\LocK}{\mathcal{K}}
\newcommand{\copyC}{\textbf{\texttt{copy}}\xspace}
\newcommand{\Trfm}{\textbf{\texttt{Trfm}}\xspace}
\newcommand{\tableT}{\textbf{\texttt{table}}\xspace}

\newcommand{\emptyVal}{\text{empty}\xspace}
\newcommand{\amount}{\text{amount}\xspace}

\newcommand{\types}{\textsc{Type}\xspace}

\newcommand{\langName}{Psamathe\xspace}
\newcommand{\langNamePronounce}{/\textipa{sAmATi}/}

\newcommand{\newc}{\textbf{\texttt{new}}}

\newcommand{\asset}{\textbf{\texttt{asset}}\xspace}
\newcommand{\fungible}{\textbf{\texttt{fungible}}\xspace}

\newcommand{\consumable}{\textbf{\texttt{consumable}}\xspace}
\newcommand{\immutable}{\textbf{\texttt{immutable}}\xspace}
\newcommand{\unique}{\textbf{\texttt{unique}}\xspace}

\newcommand{\assetT}{\texttt{asset}\xspace}

\newcommand{\immutableT}{\texttt{immutable}\xspace}

\newcommand{\dom}{\textbf{\texttt{dom}}\xspace}

\newcommand{\boolt}{\textbf{\texttt{bool}}\xspace}
\newcommand{\natt}{\textbf{\texttt{nat}}\xspace}

\newcommand{\transformer}{\textbf{\texttt{transformer}}\xspace}

\newcommand{\tryS}{\textbf{\texttt{try}}\xspace}
\newcommand{\catchS}{\textbf{\texttt{catch}}\xspace}

\newcommand{\Decl}{\textbf{\texttt{Decl}}\xspace}

\newcommand{\Prog}{\textbf{\texttt{Prog}}\xspace}
\newcommand{\Stmt}{\textbf{\texttt{Stmt}}\xspace}

\newcommand{\type}{\textbf{\texttt{type}}\xspace}
\newcommand{\is}{\textbf{\texttt{is}}\xspace}

\newcommand{\zip}{\textbf{\texttt{zip}}\xspace}

\newcommand{\emptyq}{\textbf{\texttt{empty}}\xspace}

\newcommand{\nonempty}{\textbf{\texttt{nonempty}}\xspace}
\newcommand{\any}{\textbf{\texttt{any}}\xspace}

\newcommand{\exactlyone}{\textbf{\texttt{one}}\xspace}

\newcommand{\consume}{\textbf{\texttt{consume}}\xspace}

\newcommand{\revert}{\textbf{\texttt{revert}}\xspace}
\newcommand{\error}{\textbf{\texttt{error}}\xspace}

\newcommand{\select}{\textbf{\texttt{select}}\xspace}

\newcommand{\ok}{\textbf{ok}\xspace}

\newcommand{\var}{\textbf{\texttt{var}}\xspace}

\newcommand{\flowproves}{\proves}
\newcommand{\flowprovesout}{\dashv}

\renewcommand{\id}[1]{\text{id}_{#1}}

\begin{document}

\title{\langName: A DSL with Flows for Safe Blockchain Assets}

\author{Reed Oei}
\affiliation{%
    \institution{University of Illinois}
    \city{Urbana}
    \country{USA}
}
\email{reedoei2@illinois.edu}

\author{Michael Coblenz}
\affiliation{%
    \institution{University of Maryland}
    \city{College Park}
    \country{USA}
}
\email{mcoblenz@umd.edu}

\author{Jonathan Aldrich}
\affiliation{%
    \institution{Carnegie Mellon University}
    \city{Pittsburgh}
    \country{USA}
}
\email{jonathan.aldrich@cs.cmu.edu}

\begin{abstract}
Blockchains host smart contracts for crowdfunding, tokens, and many other purposes.
Vulnerabilities in contracts are often discovered, leading to the loss of large quantities of money.
Psamathe is a new language we are designing around a new flow abstraction, reducing asset bugs and making contracts more concise than in existing languages.
We present an overview of Psamathe, including a partial formalization.
We also discuss several example contracts in Psamathe, and compare the Psamathe examples to the same contracts written in Solidity.
\end{abstract}

\maketitle

\section{Introduction}
Blockchains are increasingly used as platforms for applications called \emph{smart contracts}~\cite{Szabo97:Formalizing}, which automatically manage transactions in an mutually agreed-upon way.
Commonly proposed and implemented applications include supply chain management, healthcare, voting, crowdfunding, auctions, and more~\cite{SupplyChainUse,HealthcareUse,Elsden18:Making}.
Smart contracts often manage \emph{digital assets}, such as cryptocurrencies, or, depending on the application, bids in an auction, votes in an election, and so on.
These contracts cannot be patched after deployment, even if security vulnerabilities are discovered.
Some estimates suggest that as many as 46\% of smart contracts may have vulnerabilities~\cite{luuOyente}.
Vulnerabilities in smart contracts can lead to the loss of large quantities of money---the well-known DAO attack~\cite{DAO} caused the loss of over 40 million dollars.

\langName (\langNamePronounce) is a new programming language we are designing around \emph{flows}, which are a new abstraction representing an atomic transfer operation.
Together with features such as \emph{modifiers}, flows provide a \textbf{concise} way to write contracts that \textbf{safely} manage assets (see Section~\ref{sec:lang}).
Solidity, the most commonly-used smart contract language on the Ethereum blockchain~\cite{EthereumForDevs}, does not provide analogous support for managing assets.
Typical smart contracts are more concise in \langName than in Solidity, because \langName handles common patterns and pitfalls automatically.
A formalization of \langName is in progress~\cite{psamatheRepo}, with an \emph{executable semantics} implemented in the $\mathbb{K}$-framework~\cite{rosu-serbanuta-2010-jlap}, which is already capable of running the examples shown in Figures~\ref{fig:erc20-transfer-flow} and~\ref{fig:voting-impl-flow} (ERC-20 and a voting contract).

Other newly-proposed blockchain languages include Flint, Move, Nomos, Obsidian, and Scilla~\cite{schrans2018flint,blackshear2019move,das2019nomos,coblenz2019obsidian,sergey2019scilla}.
Scilla and Move are intermediate-level languages, whereas \langName is intended to be a high-level language.
Obsidian, Move, Nomos, and Flint use linear or affine types to manage assets; \langName uses \emph{type quantities}, which extend linear types to allow a more precise analysis of the flow of values in a program.
None of the these languages have flows or provide support for all the modifiers that \langName does.

\section{Language}\label{sec:lang}
A \langName program is made of \emph{transformers} and \emph{type declarations}.
Transformers contain \emph{flows} describing the how values are transferred between variables.
Type declarations provide a way to name types and to mark values with \emph{modifiers}, such as \flowinline{asset}.

Figure~\ref{fig:erc20-transfer-flow} shows a simple contract declaring a type and a transformer, which implements the core of ERC-20's \flowinline{transfer} function.
ERC-20 is a standard providing a bare-bones interface for token contracts managing \emph{fungible} tokens.
Fungible tokens are interchangeable (like most currencies), so it is only important how many tokens are owned by an entity, not \textbf{which} tokens.

\begin{figure}
    \centering
\begin{lstlisting}[language=flow, xleftmargin=0.3em]
type Token is fungible asset uint256
transformer transfer(balances : any map one address => any Token,
                     dst : one address, amount : any uint256) {
  balances[msg.sender] --[ amount ]-> balances[dst] |\label{line:erc20-flow-ex}|
}

\end{lstlisting}
    \caption{A \langName contract with a simple \flowinline{transfer} function, which transfers \flowinline{amount} tokens from the sender's account to the destination account.
It is implemented with a single flow, which automatically checks all the preconditions to ensure the transfer is valid.}
    \label{fig:erc20-transfer-flow}
\end{figure}

\subsection{Overview}

\langName is built around the concept of a flow.
Using the more declarative, \emph{flow-based} approach provides the following advantages over imperative state updates:
\begin{itemize}
    \item \textbf{Static safety guarantees}: Each flow is guaranteed to preserve the total amount of assets (except for flows that explicitly consume or allocate assets). 
        The total amount of a nonconsumable asset never decreases.
        Each asset has exactly one reference to it, either via a variable in the current environment, or in a table/record.
        The \flowinline{immutable} modifier prevents values from changing.
    \item \textbf{Dynamic safety guarantees}: \langName automatically inserts dynamic checks of a flow's validity; e.g., a flow of money would fail if there is not enough money in the source, or if there is too much in the destination (e.g., due to overflow).
        The \flowinline{unique} modifier, which restrict values to never be created more than once, is also checked dynamically.
    \item \textbf{Data-flow tracking}: We hypothesize that flows provide a clearer way of specifying how resources flow in the code itself, which may be less apparent using other approaches, especially in complicated contracts.
        Additionally, developers must explicitly mark when assets are \emph{consumed}, and only assets marked as \flowinline{consumable} may be consumed.
    \item \textbf{Error messages}: When a flow fails, the \langName runtime provides automatic, descriptive error messages, such as
\begin{lstlisting}[numbers=none, basicstyle=\small\ttfamily, xleftmargin=-5.0ex]
Cannot flow <amount> Token from account[<src>] to account[<dst>]:
    source only has <balance> Token.
\end{lstlisting}
        Flows enable such messages by encoding information into the source code.
\end{itemize}

Each variable and function parameter has a \emph{type quantity}, approximating the number of values, which is one of: \flowinline{empty}, \flowinline{any}, \flowinline{one}, or \flowinline{nonempty}. 
Only \flowinline{empty} asset variables may be dropped.
Type quantities are inferred if omitted; every type quantity in Figure~\ref{fig:erc20-transfer-flow} can be omitted.

\emph{Modifiers} can be used to place constraints on how values are managed: they are \flowinline{asset}, \flowinline{consumable}, \flowinline{fungible}, \flowinline{unique}, and \flowinline{immutable}.
An \flowinline{asset} is a value that must not be reused or accidentally lost, such as money.
A \flowinline{consumable} value is an \flowinline{asset} that it may be appropriate to dispose of, via the \flowinline{consume} construct, documenting that the disposal is intentional.
For example, while bids should not be lost \textbf{during} an auction, it is safe to dispose of them after the auction ends.
A \flowinline{fungible} value can be \textbf{merged}, and it is \textbf{not} \flowinline{unique}.
The modifiers \flowinline{unique} and \flowinline{immutable} provide the safety guarantees mentioned above.

We now give examples using modifiers and type quantities to guarantee additional correctness properties in the context of a lottery.
The \flowinline{unique} and \flowinline{immutable} modifiers ensure users enter the lottery at most once, while \flowinline{asset} ensures that we do not accidentally lose tickets.
We use \flowinline{consumable} because tickets no longer have any value when the lottery is over.
\begin{lstlisting}[language=flow]
type TicketOwner is unique immutable address
type Ticket is consumable asset {
    owner : TicketOwner,
    guess : uint256
}
\end{lstlisting}
Consider the code snippet in Figure~\ref{fig:lottery-end}, handling ending the lottery.
The lottery cannot end before there is a winning ticket, enforced by the \flowinline{nonempty} in the \emph{filter} on line~\ref{line:lottery-filter}; note that, as \flowinline{winners} is \flowinline{nonempty}, there cannot be a divide-by-zero error.
Without line~\ref{line:empty-lottery-balance}, \langName would give an error indicating \flowinline{balance} has type \flowinline{any ether}, not \flowinline{empty ether}---a true error, because in the case that the jackpot cannot be evenly split between the winners, there will be some ether left over.
\begin{figure}
    \centering
\begin{lstlisting}[language=flow, xleftmargin=-0.5em]
var winners : list Ticket <-- tickets[nonempty st ticketWins(winNum, _)] |\label{line:lottery-filter}|
// Split jackpot among winners
winners --> payEach(jackpot / length(winners), _)
balance --> lotteryOwner.balance |\label{line:empty-lottery-balance}|
// Lottery is over, destroy losing tickets
tickets --> consume
\end{lstlisting}
    \caption{A code snippet that handles the process of ending a lottery.}
    \label{fig:lottery-end}
\end{figure}

One could try automatically inserting dynamic checks in a language like Solidity, but in many cases it would require additional annotations.
Such a system would essentially reimplement flows, providing some benefits of \langName, but not the same static guarantees.
Some patchwork attempts already exist, such as the SafeMath library which checks for the specific case of underflow and overflow.
For example, consider the following code snippet in \langName, which performs the task of selecting a user by some predicate \flowinline{P}.
\begin{lstlisting}[language=flow]
var user : User <-- users[one such that P(_)]
\end{lstlisting}
This line expresses that we wish to select exactly one user satisfying the predicate.
There is no way to express this same constraint in Solidity (or most languages) without manually writing code to check it.
Additionally, in Solidity, variables are initialized with default values, making uniqueness difficult to enforce.


Programs in \langName are \emph{transactional}: a sequence of flows will either all succeed, or, if a single flow fails, the rest will fail as well.
If a sequence of flows fails, the error propagates, like an exception, until it either: a) reaches the top level, and the entire transaction fails; or b) reaches a \flowinline{catch}, and then only the changes made in the corresponding \flowinline{try} block will be reverted, and the code in the \flowinline{catch} block will be executed.

\section{Formalization}

We now present typing and evaluation rules for the core calculus of \langName.

\subsection{Syntax}

Figure~\ref{fig:syntax} shows the abstract syntax of the core calculus of \langName.
\begin{figure}
    \centering
    \begin{align*}
        f &\in \textsc{TransformerNames} & t &\in \textsc{TypeNames} \\
        a,x,y,z &\in \textsc{Identifiers} & n,m &\in \N \\
    \end{align*}
    \begin{tabular}{l r l l}
        $\mathcal{Q}$, $\mathcal{R}$, $\mathcal{S}$ & \bnfdef & $\exactlyone$ \bnfalt $\any$ \bnfalt $\nonempty$ \bnfalt $\emptyq$ & (type quantities) \\
        $M$ & \bnfdef & \fungible \bnfalt \unique \bnfalt \immutable & (modifiers) \\
            & \bnfalt & \consumable \bnfalt \asset & (modifiers) \\
        $T$ & \bnfdef & $\boolt$ \bnfalt $\natt$ \bnfalt $t$ \bnfalt $\tableT(\overline{x})~\tau$ \bnfalt $\curlys{ \overline{x : \tau} }$ & (base types) \\
        $\tau$, $\sigma$, $\pi$ & \bnfdef & $\mathcal{Q}~T$ & (types) \\
        $\Loc$, $\LocK$ & \bnfdef & $\true$ \bnfalt $\false$ \bnfalt $n$ & \\
               & \bnfalt & $x$ \bnfalt $\Loc.x$ \bnfalt $\var~x : T$ \bnfalt $[ \overline{\Loc} ]$ \bnfalt $\{ \overline{x : \tau \mapsto \Loc} \}$ & \\
               & \bnfalt & $\copyC(\Loc)$ \bnfalt $\zip(\overline{\Loc})$ & \\
               & \bnfalt & $\Loc[\Loc]$ \bnfalt $\Loc[\mathcal{Q} \tsuchthat f(\overline{\Loc})]$ \bnfalt $\consume$ & \\
        $\Trfm$ & \bnfdef & $\newc~t(\overline{\Loc})$ \bnfalt $f(\overline{\Loc})$ & (transformer calls) \\
        $\Stmt$ & \bnfdef & $\Loc \to \Loc$ \bnfalt $\Loc \to \Trfm \to \Loc$ & (flows) \\
                & \bnfalt & $\tryS~\{ \overline{\Stmt} \}~\catchS~\{ \overline{\Stmt} \}$ & (try-catch) \\
        $\Decl$ & \bnfdef & $\transformer~f(\overline{x : \tau}) \to x : \tau ~\{ \overline{\Stmt} \}$ & (transformers) \\
                & \bnfalt & $\type~t~\is~\overline{M}~T$ & (type decl.) \\
        $\Prog$ & \bnfdef & $\overline{\Decl}; \overline{\Stmt}$ & (programs)
    \end{tabular}
    \caption{Syntax of the core calculus of \langName.}
    \label{fig:syntax}
\end{figure}

\subsection{Statics}
Below we show the type rules needed to check flows between variables.
We use $\Gamma$ and $\Delta$ as type environments, pairs of variables and types, identified with partial functions between the two.

First, we discuss rules checking the types of the source and destination locators of a flow.
\emph{Locators} are expressions that evaluate to \emph{locations}, identifying \emph{resources} in the program storage.
The simplest example of a locator is just a variable $x$, which evaluates to the location that stores the values of $x$.
We can build up more complicated locators using some familiar operations, such as a field accesses, $x.f$, which evaluates to the location of the field $f$ inside of $x$, or $x[y]$, which evaluates to the location the value specified by $y$ in $x$ (e.g., if $x$ is a map, then the value whose key is $y$).
To properly type locators, we must consider how the locator is being used---locators used as sources will decrease in quantity, whereas locators used as destinations will increase in quantity; after a locator is used, we must update the type of the located values in the type environment.
This is all captured in the following judgement.

\framebox{$\Gamma \flowproves_M f ; \Loc : \tau \flowprovesout \Delta $} \textbf{Locator Typing}
This judgement states in the environment $\Gamma$ and \emph{mode} $M$, using $\Loc$ according to the \emph{updater} $f$ will yield a value of type $\tau$ and the new type environment $\Delta$.

A \emph{mode} $M$ is either $S$, meaning source, or $D$, meaning destination.
This ensures that we don't use, for example, numeric literals as the destination of a flow.
We refer to $\Gamma$ as the \emph{input environment} and $\Delta$ as the \emph{output environment}.
We use $f$ to refer to a function on types ($\types \to \types$), called an \emph{updater}.
We call such functions \emph{updaters}.
We adopt the convention that if $f : \types^n \to \types$ and $g_1,\ldots,g_n : \types^m \to \types$, then $f(g_1, \ldots, g_n) : \types^m \to \types$, where
\[
    f(g_1, \ldots, g_n)(\overline{\tau}) = f(g_1(\overline{\tau}), \ldots, g_n(\overline{\tau}))
\]
We use the following type functions:
\begin{itemize}
    \item $\id{\types}$ is the identity function on types
    \item $\oplus \mathcal{Q}, \ominus \mathcal{Q} : \types \to \types$ are functions that take a type $\tau$ and add/subtract $\mathcal{Q}$ to/from its type quantity.
    \item $\with_{\mathcal{Q}} : \types \to \types$ is the function that replaces the type quantity of $\tau$ with $\mathcal{Q}$; i.e., $\with_{\mathcal{Q}}(\mathcal{Q}'~T) = \mathcal{Q}~T$
    \item $\max : \types^2 \to \types$ returns the type with the larger type quantity
    \item $\sqcup : \types^2 \to \types$ performs the \emph{join} of the two type quantities according to specificity (e.g., $\emptyq \sqcup \nonempty = \any$)
\end{itemize}

Note that we write $\Gamma \flowproves_M \overline{f ; \Loc : \tau} \flowprovesout \Delta$ where $|\overline{\Loc}| = n$ to mean
``for all $1 \leq i \leq n$, $\Gamma_i \flowproves_M f_i ; \Loc_i : \tau_i \flowprovesout \Delta_i$'' where $\Gamma = \Gamma_1$, $\Gamma_{i+1} = \Delta_{i}$, and $\Delta = \Delta_n$.

Constants of type $\natt$ or $\boolt$ can only be used as sources---it doesn't make sense to flow values to a constant.
We define $\# : \N \to \textsc{TypeQuant}$ so that $\#(n)$ is the best approximation by type quantity of $n$, i.e.,
\[
    \#(n) =
    \begin{cases}
        \emptyq & \tif n = 0 \\
        \exactlyone & \tif n = 1 \\
        \nonempty & \tif n > 1 \\
    \end{cases}
\]

\begin{mathpar}
    \inferrule*[right=Nat]{
    }{ \Gamma \flowproves_S f ; n : \#(n)~\natt \flowprovesout \Gamma }

    \inferrule*[right=Bool]{
        b \in \curlys{ \true, \false }
    }{ \Gamma \flowproves_S f ; b : \exactlyone~\boolt \flowprovesout \Gamma }
\end{mathpar}

Variables may be used as either sources or destinations, as long as they are not immutable.
We may also use them to select resources (in which case $f = \id{\types}$), even if immutable.
\begin{mathpar}
    \inferrule*[right=Var]{
        \tau~\immutableT \implies f = \id{\types}
    }{ \Gamma, x : \tau \flowproves_M f ; x : \tau \flowprovesout \Gamma, x : f(\tau) }
\end{mathpar}

Variable definitions must be destinations, as newly defined variables are always empty, so there is no reason to use them as sources.
\begin{mathpar}
    \inferrule*[right=VarDef]{
    }{ \Gamma \flowproves_D f ; (\var~x : T) : \emptyq~T \flowprovesout \Gamma, x : f(\emptyq~T) }
\end{mathpar}

We now consider type-checking statements.

\framebox{$\Gamma \flowproves S~\ok \flowprovesout \Delta$} \textbf{Statement Well-formedness}
This judgement states that in the environment $\Gamma$, the statement $S$ is well-formed and it transforms $\Gamma$ into the output environment $\Delta$.

To check that a flow is well-formed, we check that $\Loc$ and $\LocK$ have the same base type.
We use $\with_{\emptyq}$ to clear all the that $\Loc$ locates, and we use $(\oplus \mathcal{Q})$ to add these to $\LocK$.
\begin{mathpar}
    \inferrule*[right=Ok-Flow]{
        \Gamma \flowproves_S \with_{\emptyq} ; \Loc : \mathcal{Q}~T \flowprovesout \Delta
        \\
        \Delta \flowproves_D (\oplus \mathcal{Q}) ; \LocK : \mathcal{R}~T ; \flowprovesout \Xi
    }{ \Gamma \flowproves (\Loc \to \LocK)~\ok \flowprovesout \Xi }
\end{mathpar}


\framebox{$\flowproves \Decl~\ok$} \textbf{Declaration Well-formedness}
This judgement states that a declaration is well-formed.


To check that a transformer declaration is well-formed, we must check the behavior of its body: it must not leave any assets unused, the \emph{return variable}, $z$ must be of the correct type, and the \emph{auxiliary arguments}, $\overline{x : \tau}$, must have the same type as they did at the beginning of the transformer.
The latter requirement ensures that transformers may be repeatedly called; for example, if used as predicate to filter a list with many elements.
Note that $y$ may be unused, if it is not an asset.
\begin{mathpar}
    \inferrule*[right=Ok-Transformer]{
        \overline{x : \tau}, y : \tau_y, z : \with_{\emptyq}(\tau_z) \flowproves \overline{\Stmt~\ok} \flowprovesout \Delta, \overline{x : \tau}, z : \tau_z
        \\
        \forall v : \sigma \in \Delta. \lnot(\sigma~\assetT)
    }{ (\transformer f(\overline{x : \tau}, y : \tau_y) \to z : \tau_z~\{ \overline{\Stmt} \}) ~\ok }
\end{mathpar}

\subsection{Dynamics}
Below are the rules to evaluate statements of flows between variables.

We introduce sorts for \emph{values}, \emph{resources}, values tagged with their type, and storage values.
Storage values are either a natural number, indicating a location in the store, or $\amount(n)$, indicating $n$ of some resource.
Locators evaluate to storage value pairs, i.e., $(\ell, k)$, where $\ell$ indicates the parent location of the value, and $k$ indicates which value to select from the parent location.
If $\ell = k$, then every value should be selected.
This is useful because it allows us to locate only part of a fungible resources, or a specific element inside a list.
The $\select(\rho, \ell, k)$ construct resolves storage value pairs into the resource that should be selected.

\begin{tabular}{l r l l}
    $V$ & \bnfdef & $n$ \bnfalt $\error$ & (values) \\
    $R$ & \bnfdef & $(T, V)$ & (resources) \\
    $\ell, k$ & \bnfdef & $n$ \bnfalt $\amount(n)$ & (storage values) \\
    $\Loc$ & \bnfdef & $\ldots$ \bnfalt $(n, \ell)$ \\
    $\Stmt$ & \bnfdef & $\ldots$ \bnfalt $\revert$ \\
\end{tabular}

We accordingly expand type environments to contain $(n, \ell)$ pairs.

\begin{definition}
    A (runtime) environment $\Sigma$ is a tuple $(\mu, \rho)$ where $\mu : \textsc{IdentifierNames} \partialfunc \mathbb{N} \times \ell$ is the \emph{variable lookup environment}, and $\rho : \mathbb{N} \partialfunc R$ is the \emph{storage environment}.
\end{definition}

We now give rules for how to evaluate programs in \langName.
We begin with rules to evaluate locators.

\framebox{$\angles{\Sigma, \Loc} \to \angles{\Sigma', \Loc'}$} \textbf{Locator Evaluation}
This judgement states that with the environment $\Sigma$, $\Loc$ steps to $\Loc'$ and updates the environment to $\Sigma'$.

Note that $(\ell, \amount(n))$ and $(\ell, \ell)$ are equivalent w.r.t. $\select$ when $\rho(\ell) = (T, n)$ for some fungible $T$.

\begin{mathpar}
    \inferrule*[right=Loc-Nat]{
        m \not\in \dom(\rho)
    }{ \angles{ (\mu, \rho), n } \to \angles{ (\mu, \rho[m \mapsto (\natt, n)]), (m, \amount(n))) } }

    \inferrule*[right=Loc-Bool]{
        n \not\in \dom(\rho)
    }{ \angles{ (\mu, \rho), b } \to \angles{ (\mu, \rho[n\mapsto (\boolt, b)]), (n, n) } }

    \inferrule*[right=Loc-Id]{
        \mu(x) \neq \bot
    }{ \angles{ \Sigma, x } \to \angles{ \Sigma, \mu(x) } }

    \inferrule*[right=Loc-VarDef]{
        n \not\in \dom(\rho)
    }{ \angles{ \Sigma, \var~x : T } \to \angles{ (\mu[x \mapsto n], \rho[n \mapsto \emptyVal(T)]), (n, n) } }

\end{mathpar}

Next, the rule for evaluating statements.

\framebox{$\angles{\Sigma, \overline{\Stmt}} \to \angles{\Sigma', \overline{\Stmt'}}$} \textbf{Statement Evaluation}

Thsi judgement states that in the environmen $\Sigma$, the statements $\overline{\Stmt}$ step to $\overline{\Stmt'}$, and update the environment to $\Sigma'$.
Note that when the list of statements is empty, we omit it; that is, we write $\angles{\Sigma, \overline{\Stmt}} \to \Sigma'$, not $\angles{\Sigma, \overline{\Stmt}} \to \angles{\Sigma', \cdot}$.

To evaluate a flow, we must resolve the selected resources, subtract them from their parent locations, and finally add them all to the destination location.
If either the subtraction or addition results in an error, the whole flow causes a \revert.

\begin{mathpar}
    \inferrule*[right=Flow]{
        \select(\rho, n, \ell) = R
        \\
        \error \not\in \curlys{R, \rho(m) + R}
    }{ \angles{ (\mu, \rho), (n, \ell) \to (m, k) } \to (\mu, \rho[n \mapsto \rho(n) - R, m \mapsto \rho(m) + R]) }

    \inferrule*[right=Flow-Error]{
        \select(\rho, n, \ell) = R
        \\
        \error \in \curlys{R, \rho(m) + R}
    }{ \angles{ \Sigma, (n, \ell) \to (m, k) } \to \angles{ \Sigma, \revert } }
\end{mathpar}

\section{Examples}

In this section, we present additional examples, showing that \langName and flows are useful for a variety of smart contracts.
We also show examples of these same contracts in Solidity, and compare the \langName implementations to those in Solidity.

\subsection{ERC-20 in Solidity}\label{sec:erc20-impl}
Each ERC-20 contract manages the ``bank accounts'' for its own tokens, keeping track of how many tokens each account has; accounts are identified by addresses.
We compare the \langName implementation in Figure~\ref{fig:erc20-transfer-flow} to Figure~\ref{fig:erc20-transfer-sol}, which shows a Solidity implementation of the same function.
In this case, the sender's balance must be at least as large as \flowinline{amount}, and the destination's balance must not overflow when it receives the tokens.
\langName automatically inserts code checking these two conditions, ensuring the checks are not forgotten.
As noted above, we can automatically generate descriptive error messages with no additional code, which are not present in the Solidity implementation.
\begin{figure}
    \centering
\begin{lstlisting}[language=Solidity, xleftmargin=-0.5em]
mapping (address => uint256) balances;
function transfer(address dst, uint256 amount) public {
  require(amount <= balances[msg.sender]);
  balances[msg.sender] = balances[msg.sender].sub(amount);
  balances[dst] = balances[dst].add(amount);
}

\end{lstlisting}
    \caption{An implementation of ERC-20's \lstinline[language=Solidity]{transfer} function in Solidity from one of the reference implementations~\cite{erc20Consensys}.
        All preconditions are checked manually.
        Note that we must include the \lstinline[language=Solidity]{SafeMath} library (not shown) to use the \lstinline[language=Solidity]{add} and \lstinline[language=Solidity]{sub} functions, which check for underflow/overflow.}
    \label{fig:erc20-transfer-sol}
\end{figure}

\subsection{Voting}\label{sec:voting-impl}
One proposed use for blockchains is for voting~\cite{Elsden18:Making}.
Figure~\ref{fig:voting-impl-flow} shows the core of an implementation of a voting contract in \langName.
Each contract instance has several proposals, and users must be given permission to vote by the chairperson, assigned in the constructor of the contract (not shown).
Each eligible voter can vote exactly once for exactly one proposal, and the proposal with the most votes wins.
This example shows some uses of the $\unique$ modifier; in this contract, $\unique$ ensures that each user, represented by an \lstinline{address}, can be given permission to vote at most once, while the use of $\asset$ ensures that votes are not lost or double-counted.
This example show that \langName, as well as flows, are suited to a wide range of common smart contract applications.
\begin{figure}
    \centering
\begin{lstlisting}[language=flow, xleftmargin=-1em]
type Voter is unique immutable asset address
type ProposalName is unique immutable asset string
type Election is asset {
  chairperson : address,
  eligibleVoters : set Voter,
  proposals : map ProposalName => set Voter
}
transformer giveRightToVote(this : Election, voter : address) {
  only when msg.sender = this.chairperson
  new Voter(voter) --> this.eligibleVoters
}
transformer vote(this : Election, proposal : string) {
  this.eligibleVoters --[ msg.sender ]-> this.proposals[proposal]
}

\end{lstlisting}
    \caption{A simple voting contract in \langName.}
    \label{fig:voting-impl-flow}
\end{figure}

Figure~\ref{fig:voting-impl-sol} shows an implementation of the same voting contract in Solidity, based on the Solidity by Example tutorial~\cite{solidityByExample}.
Again, we must manually check all preconditions.
\begin{figure}
    \centering
\begin{lstlisting}[language=Solidity, xleftmargin=-1em]
contract Ballot {
  struct Voter { uint weight; bool voted; uint vote; }
  struct Proposal { bytes32 name; uint voteCount; }
  address public chairperson;
  mapping(address => Voter) public voters;
  Proposal[] public proposals;
  function giveRightToVote(address voter) public {
    require(msg.sender == chairperson,
      "Only chairperson can give right to vote.");
    require(!voters[voter].voted, "The voter already voted.");
    voters[voter].weight = 1;
  }
  function vote(uint proposal) public {
    Voter storage sender = voters[msg.sender];
    require(sender.weight != 0, "No right to vote");
    require(!sender.voted, "Already voted.");
    sender.voted = true;
    sender.vote = proposal;
    proposals[proposal].voteCount += sender.weight;
  }
}

\end{lstlisting}
    \caption{A simple voting contract in Solidity.}
    \label{fig:voting-impl-sol}
\end{figure}

\subsection{Blind Auction}\label{sec:blind-auction-impl}
Another proposed use of blockchains is auctions~\cite{Elsden18:Making}.
Figure~\ref{fig:blind-auction-impl-flow} shows an implementation of the \emph{reveal phase} of a \emph{blind auction} in \langName.
A blind auction is an auction in which bids are placed, but not revealed until the auction has ended, meaning that other bidders have no way of knowing what bids have been placed so far.
Because transactions on the Ethereum blockchain are publicly viewable, the bids must be blinded cryptographically, in this case, using the KECCAK-256 algorithm~\cite{bertoni2013keccak}.
Bidders sent the hashed bytes of their bid, that is, the value (in ether) and some secret string of bytes, along with a deposit of ether, which must be at least as large as the intended value of the bid for the bid to be valid.
After bidding is over, they must \emph{reveal} their bid by sending a transaction containing these details, which will be checked by the \flowinline{Auction} contract (line~\ref{line:auction-check-bid}).
Any extra value in the bid (used to mask the true value of the bid), will be returned to the bidder.

This example uses a pipeline of locators and transformers (lines~\ref{line:auction-pipeline-start}-\ref{line:auction-pipeline-end}) to concisely process each revealed bid, showing another case in which flows provide a clean way to write smart contracts.

\begin{figure}
    \centering
\begin{lstlisting}[language=flow]
type Bid is consumable asset {
  sender : address,
  blindedBid : bytes,
  deposit : ether
}
type Reveal is { value : nat, secret : bytes }
type Auction is asset {
  biddingEnd : nat, revealEnd : nat, ended : bool,
  bids : map address => list Bid,
  highestBidder : address, highestBid : ether,
  pendingReturns : map address => ether
}
transformer reveal(this : Auction, reveals : list Reveal) {
  only when biddingEnd <= now and now <= revealEnd
  zip(this.bids[msg.sender], reveals) |\label{line:auction-pipeline-start}|
    --[ any such that _.fst.blindedBid = keccak256(_.snd) ] |\label{line:auction-check-bid}|
    --> this.revealBid(_.fst, _.snd) |\label{line:auction-pipeline-end}|
}
transformer revealBid(this : Auction, bid : Bid, reveal : Reveal) {
  try {
    only when reveal.value >= this.highestBid
    this.highestBid --> this.pendingReturns[highestBidder]
    bid.deposit --[ reveal.value ]-> this.highestBid
    bid.sender --> this.highestBidder
  } catch {}
  bid.deposit --> bid.sender.balance
  bid --> consume
}
\end{lstlisting}
    \caption{Implementation of reveal phase of a blind auction contract in \langName.}
    \label{fig:blind-auction-impl-flow}
\end{figure}

\section{Conclusion and Future Work}

We have presented the \langName language for writing safer smart contracts.
\langName uses the new flow abstraction, assets, and type quantities to provide its safety guarantees.
We have shown example smart contracts in both \langName and Solidity, showing that \langName is capable of expressing common smart contract functionality in a concise manner, while retaining key safety properties.

In the future, we plan to fully implement the \langName language, and prove its safety properties.
We also hope to study the benefits and costs of the language via case studies, performance evaluation, and the application of flows to other domains.
Finally, we would also like to conduct a user study to evaluate the usability of the flow abstraction and the design of the language, and to compare it to Solidity, which we hypothesize will show that developers write contracts with fewer asset management errors in \langName than in Solidity.

\bibliographystyle{ACM-Reference-Format}
\bibliography{biblio}

\end{document}